**Generation of Terahertz Radiation by Wave Mixing in Zigzag Carbon Nanotubes**

S.Y.Mensah[a], S. S. Abukari[a], N. G. Mensah[b], K. A. Dompreh[a], A. Twum[a] and F. K. A. Allotey[c]

[a]*Department of Physics, Laser and Fibre Optics Centre, University of Cape Coast, Cape Coast, Ghana*

[b]*Department of Mathematics, University of Cape Coast, Cape Coast, Ghana*

[c]*Institute of Mathematical Sciences, Accra, Ghana*

[*]*Corresponding author.* [a]*Department of Physics, Laser and Fibre Optics Centre, University of Cape Coast, Cape Coast, Ghana*

Tel.:+233 042 33837

*E-mail address*: profsymensah@yahoo.co.uk

(S. Y. Mensah)

**Abstract**

With the use of the semiclassical Boltzmann equation we have calculated a direct current (d.c) in undoped zigzag carbon nanotube (CN) by mixing two coherent electromagnetic waves with commensurate frequencies i.e $\omega_1 = \Omega$ and $\omega_2 = 2\Omega$. This effect is attributed to the nonparabolicity of the electron energy band which is very strong in carbon nanotubes. We observed that the current $j$ is negative similar to that observed in superlattice. However if the phase shift $\varphi$ lies between $\frac{\pi}{2}$ and $\frac{3\pi}{2}$ there is an inversion and the current becomes positive. It is interesting to note that $j$ exhibit negative differential conductivity as expected for d.c through carbon nanotubes. This method can be used to generate terahertz radiation in carbon nanotubes. It can also be used in determining the relaxation time of electrons in carbon nanotubes.

1. **Introduction**

It is a known fact that coherent mixing of waves with commensurate frequencies in a nonlinear medium can result in a product which has a zero frequency or static (d.c) electromagnetic field. If such a nonlinear interference phenomenon happens in a semiconductor or semiconductor device, then the static electric field may result into a d.c current or a dc voltage generation [1].

Infact, several mechanisms of nonlinearity could be responsible for the wave mixing in semiconductors [2-4]. Important among them is the heating mechanism where the nonlinearity is related to the dependence of the relaxation constant on the electric field [4-7]. Goychuk and Hänggi [8] have also suggested another scheme of quantum rectification using wave mixing of an alternating electric field and its second harmonic in a single miniband superlattice (SL). Their approach is based on the theory of quantum ratchets and therefore the necessary conditions for the appearance of dc include a dissipation (quantum noise) and an extended periodic system [8].

Interesting to this paper is where the mechanism of nonlinearity is due to the nonparabolocity of the electron energy spectrum. Notable among such materials are the superlattice (SL) and carbon nanotubes (CNs). In superlattice the theory of wave mixing based on a solution of the Boltzmann equation have been studied in [9-11]. In all these works, the situation where $E(t) = E_1 cos\Omega t + E_2 cos(2\Omega t + \varphi)$ were not studied directly. The first paper to study this situation in SL can be found in [12]. Recently this problem has been revisited in the following papers [1,13, 14] because of the interest it generates. We study this effect in zigzag carbon nanotubes.

This work will be organised as follows: section 1 deals with introduction; in section 2, we establish the theory and solution of the problem; section 3, we discuss the results and draw conclusion.

## 2. Theory

Following the approach of [15] we consider an undoped single-wall zigzag (n, 0) carbon nanotubes (CNs) subjected to the electric mixing harmonic fields.

$$E(t) = E_1 \cos\omega_1 t + E_2 \cos(\omega_2 t + \theta) \tag{1}$$

We further consider the semiclassical approximation in which the motion of $\pi$-electrons are considered as classical motion of free quasi-particles in the field of crystalline lattice with dispersion law extracted from the quantum theory.

Considering the hexagonal crystalline structure of CNs and the tight binding approximation, the dispersion relation is given as

$$\varepsilon(s\Delta p_\varphi, p_z) \equiv \varepsilon_s(p_z) = \pm\gamma_0 \left[1 + 4\cos(ap_z)\cos\left(\frac{a}{\sqrt{3}}s\Delta p_\varphi\right) + 4\cos^2\left(\frac{a}{\sqrt{3}}s\Delta p_\varphi\right)\right]^{1/2} \tag{2}$$

for zigzag CNs [15]

Where $\gamma_0 \sim 3.0 eV$ is the overlapping integral, $p_z$ is the axial component of quasimomentum, $\Delta p_\varphi$ is transverse quasimomentum level spacing and $s$ is an integer. The expression for $a$ in Eq (2) is given as

$$a = 3a_{c-c}/2\hbar \tag{3}$$

Where $a_{c-c} = 0.142 nm$ is the C-C bond length and $\hbar$ is Plank's constant divided by $2\pi$. The - and + signs correspond to the valence and conduction bands, respectively. Due to the transverse quantization of the quasi-momentum, its transverse component can take $n$ discrete values, $p_\varphi = s\Delta p_\varphi = \pi\sqrt{3}\,s/an$ $(s = 1....,n)$

Unlike transverse quasimomentum $p_\varphi$, the axial quasimomentum $P_z$ is assumed to vary continuously within the range $0 \leq p_z \leq 2\pi/a$, which corresponds to the model of infinitely long CN $(L = \infty)$. This model is applicable to the case under consideration because we are restricted to temperatures and /or voltages well above the level spacing [16], ie. $k_B T \gg \varepsilon_C, \Delta\varepsilon$, where $k_B$ is Boltzmann constant, $T$ is the temperature, $\varepsilon_C$ is the charging energy. The energy level spacing $\Delta\varepsilon$ is given by

$$\Delta\varepsilon = \pi\hbar v_F/L \tag{4}$$

where $v_F$ is the Fermi speed and L is the carbon nanotube length [17].

Employing Boltzmann equation with a single relaxation time approximation.

$$\frac{\partial f(p)}{\partial t} + eE(t)\frac{\partial f(p)}{\partial P} = -\frac{[f(p) - f_0(p)]}{\tau} \tag{5}$$

Where e is the electron charge, $f_0(p)$ is the equilibrium distribution function, $f(p)$

is the distribution function, and $\tau$ is the relaxation time. The electric field $E(t)$ is applied along CNs axis. The relaxation term of Eq (5) describes the electron-phonon scattering [18, 19] electron-electron collisions, etc.

Expanding the distribution functions of interest in Fourier series as;

$$f_o(p) = \Delta p_\varphi \sum_{s=1}^{n} \delta(p_\varphi - s\Delta p_\varphi) \sum_{r\neq 1} f_{rs} e^{iarp_z} \quad (6)$$

and

$$f(p,t) = \Delta p_\varphi \sum_{s=1}^{n} \delta(p_\varphi - s\Delta p_\varphi) \sum_{r\neq 1} f_{rs} e^{iarp_z} \emptyset_v(t) \quad (7)$$

Where the coefficient, $\delta(x)$ is the Dirac delta function, $f_{rs}$ is the coefficient of the Fourier series and $\emptyset_v(t)$ is the factor by which the Fourier transform of the nonequilibrium distribution function differs from its equilibrium distribution counterpart.

$$f_{rs} = \frac{a}{2\pi \Delta p_\varphi s} \int_0^{\frac{2\pi}{a}} \frac{e^{-ibrp_z}}{1 + exp(\varepsilon_s(p_z)/k_B T)} dp_z \quad (8)$$

Substituting Eqs. (6) and (7) into Eq. (5), and solving with Eq. (1) we obtain

$$\emptyset_v(t) = \sum_{k_1,k_2=-\infty}^{\infty} \sum_{v_1,v_2=-\infty}^{\infty} J_{k_1}(\beta_1) J_{k_2}(\beta_2) J_{k_1+v_1}(\beta_1) J_{k_2+v_2}(\beta_2)$$

$$\times \left(\frac{(1 - i(k_1\omega_1 + k_2\omega_2)\tau)}{1 + ((k_1\omega_1 + k_2\omega_2)\tau)^2}\right) \times \{cos(v_1\omega_1 t + v_2(\omega_2 t + \theta))$$
$$- isin(v_1\omega_1 t + v_2(\omega_2 t + \theta))\} \quad (9)$$

where $\beta_1 = \frac{earE_1}{\omega_1}$, $\beta_2 = \frac{earE_2}{\omega_2}$, and $J_k(\beta)$ is the Bessel function of the k$^{th}$ order.

Similarly, expanding $\varepsilon_s(p_z)/\gamma_0$ in Fourier series with coefficients $\varepsilon_{rs}$

$$\frac{\varepsilon_s(p_s, s\Delta p_\varphi)}{\gamma_0} = \varepsilon_s(p_z) = \sum_{r\neq 0} \varepsilon_{rs} e^{igarp_z} \quad (10)$$

Where $\varepsilon_{rs} = \frac{a}{2\pi \gamma_0} \int_0^{\frac{2\pi}{a}} \varepsilon_s(p_z) e^{-iarp_z} dp_z$  (11)

and expressing the velocity as
$$v_z(p_z, s\Delta p_\varphi) = \frac{\partial \varepsilon_s(p_z)}{\partial p_z} = \gamma_0 \sum_{r\neq 0} iar \, \varepsilon_{rs} e^{iarp_z}$$  (12)

We determine the surface current density as
$$j_z = \frac{2e}{(2\pi\hbar)^2} \iint f(P) v_z(P) d^2p$$

or
$$j_z = \frac{2e}{(2\pi\hbar)^2} \sum_{s=1}^{n} \int_0^{\frac{2\pi}{a}} f(P_z, s\Delta p_\varphi) \emptyset_v(t) v_z(P_z, s\Delta p_\varphi) dp_z$$  (13)

and the integration is taken over the first Brillouin zone. Substituting Eqs. (7), (9) and (12) into (13) and linearizing with respect to $E_2$ using $J_{\pm 1}(\beta_2) \sim \beta_2/2$ ; $J_0(\beta_2) \sim 1 - \left(\beta_2/4\right)$ and then averaging the result with respect to time $t$, we obtain the direct current subjected to $\omega_1 = \Omega$ and $\omega_2 = 2\Omega$ as follows;

$$j_z = \frac{2e^2 \gamma_0 a}{\sqrt{3}\, \hbar n \alpha_{c-c}} E_2 \cos\varphi \sum_{r=1}^{\infty} r^2 \sum_{k=-\infty}^{\infty} \frac{k J_k(\beta_1) J_{k-2}(\beta_1)}{1 + (k\Omega\tau)^2} \sum_{s=1}^{n} f_{rs} \varepsilon_{rs}$$  (14)

Subsequently $\Omega\tau$ will be represented by $z_c$.

3. Results, Discussion and Conclusion

Using the solution of the Boltzmann equation with constant relaxation time τ, the exact expression for current density in CNs subjected to an electric field with two frequencies $\omega_1 = \Omega$ and $\omega_2 = 2\Omega$ was obtained after cumbersome analytical manipulation.

We noted that the current density $j_z$ is dependent on the electric field $E_2$ and $E_1$, the phase difference $\varphi$, the frequency $\Omega$, the relaxation time $\tau$ and $n$. To further understand how these parameters affect $j_z$, we sketched equation (14) using Matlab. Fig.1 represents the graph of $j_z/j_0$ on $\beta_1$ for $z_c = 0.3, 0.5, 0.9, 1$ and $2$. We observed that the current decreases rapidly, reaches a minimum value, $\beta_{min}$ and rises. For $z_c \ll 1$, the current density rises monotonously while for $z_c \gtrsim 1$, the current rises and then oscillates. This indicates that at low frequency there is rectification while as at high frequency some fluctuations occur. The rectification can be attributed to non ohmicity of the carbon nanotube for the situation where it Bloch oscillates. The behaviour of the current is similar to that observed in SL [12]. See Fig. 2. In comparison with the result in [12] for $z_c = 1$ the ratio $\left|\frac{j_{min}^{CNs}}{j_{min}^{SL}}\right| \approx 33$ which is quite substantial. We also observed a shift of the $j_{min}$ to the left with increasing value of $z_c$.

We sketched also the graph of $j_z/j_0$ against $z_c$ for $\beta_1 = 0.3, 0.5, 0.9, 1\ and\ 2$. The graph also displayed a negative differential conductivity. See Fig. 3. Interestingly like in SL as indicated in [1] the current is always positive and has a maximum at the value $z_{c\ max} \approx 0.71$ irrespective of the amplitude of the electric field $E_1$. It is worthwhile to note that $z_{c\ max}$ can be used to determine the relaxation time of the electrons in the nanotube. e.g. $\tau \approx \frac{0.71}{\Omega}$ so knowing $\Omega$ you can determine $\tau$. On the other hand for typical value for $\tau$ of $10^{-13}s$ the frequency $\frac{\Omega}{2\pi}$ would be 1.2 THz.

Finally we sketched a 3 dimensional graph of the current against $\beta_1$ and $n$. See Fig .4. It is important to note that when the phase shift $\varphi$ lies between $\frac{\pi}{2}$ and $\frac{3\pi}{2}$ there is an inversion. See Fig. 5.

In conclusion, we have studied the direct current generation due to the harmonic wave mixing in zigzag carbon nanotubes and suggest the use of this approach in generation of THz radiation .The experimental conditions for an observation of the dc current effect are practically identical to those fulfilled in a recent experiment on the generation of harmonics of the THz radiation in a semiconductor superlattice [20]. This method can also be used to determine the relaxation time $\tau$.

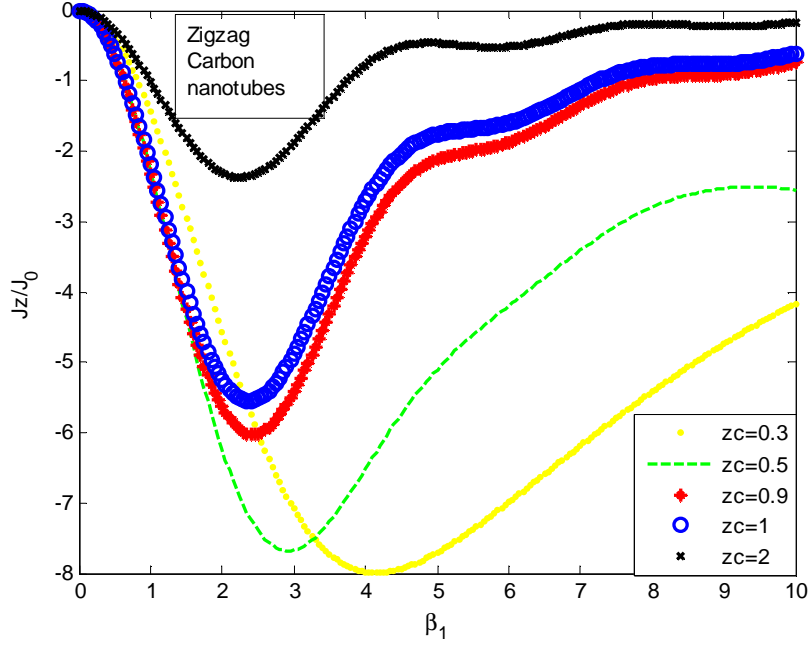

**Fig. 1.** $J_z/J_0$ is plotted against $\beta_1$ for (‐ ‐) $zc = \Omega\tau = 0.3$; (⋯) $zc = \Omega\tau = 0.5$; (∗∗∗) $zc = \Omega\tau = 0.9$ (ᵒᵒᵒ) $zc = \Omega\tau = 1$; (⋯) $zc = \Omega\tau = 2$.

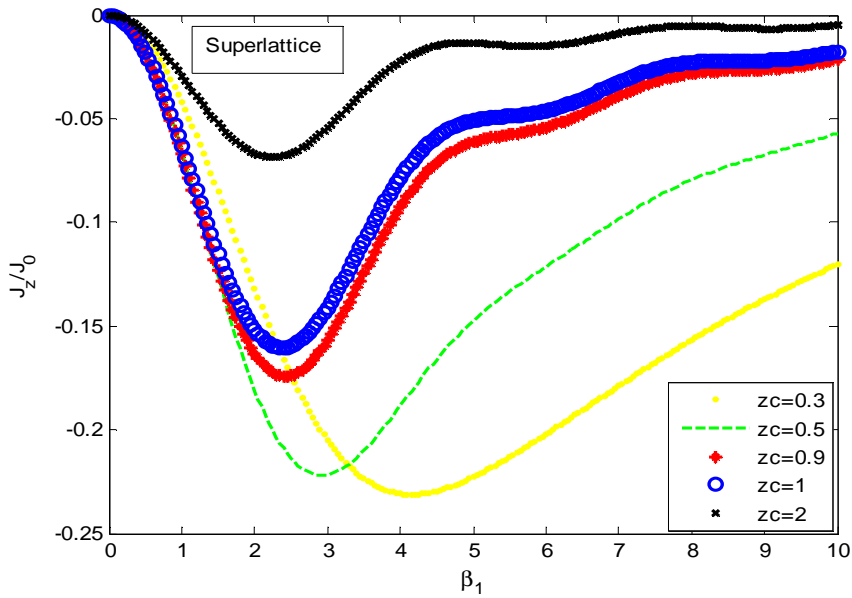

**Fig. 2.** $J_z/J_0$ is plotted against $\beta_1$ for (‐ ‐) $zc = \Omega\tau = 0.3$; (⋯) $zc = \Omega\tau = 0.5$; (∗∗∗) $zc = \Omega\tau = 0.9$ (ᵒᵒᵒ) $zc = \Omega\tau = 1$; (⋯) $zc = \Omega\tau = 2$.

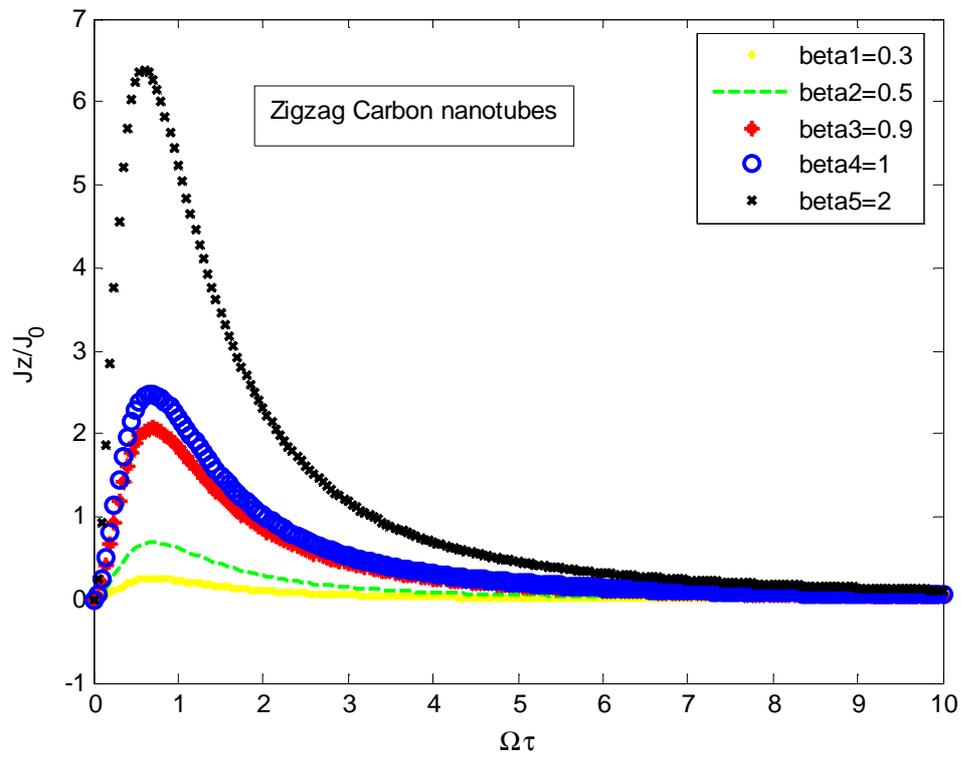

**Fig. 3.** $J_z/J_0$ is plotted against $zc = \Omega\tau$ for (– – –) $\beta_1 = 0.3$; (- - -) $\beta_1 = 0.5$; (∗∗∗) $\beta_1 = 0.9$ ; (ooo) $\beta_1 = 1$; (⋯) $\beta_1 = 2$.

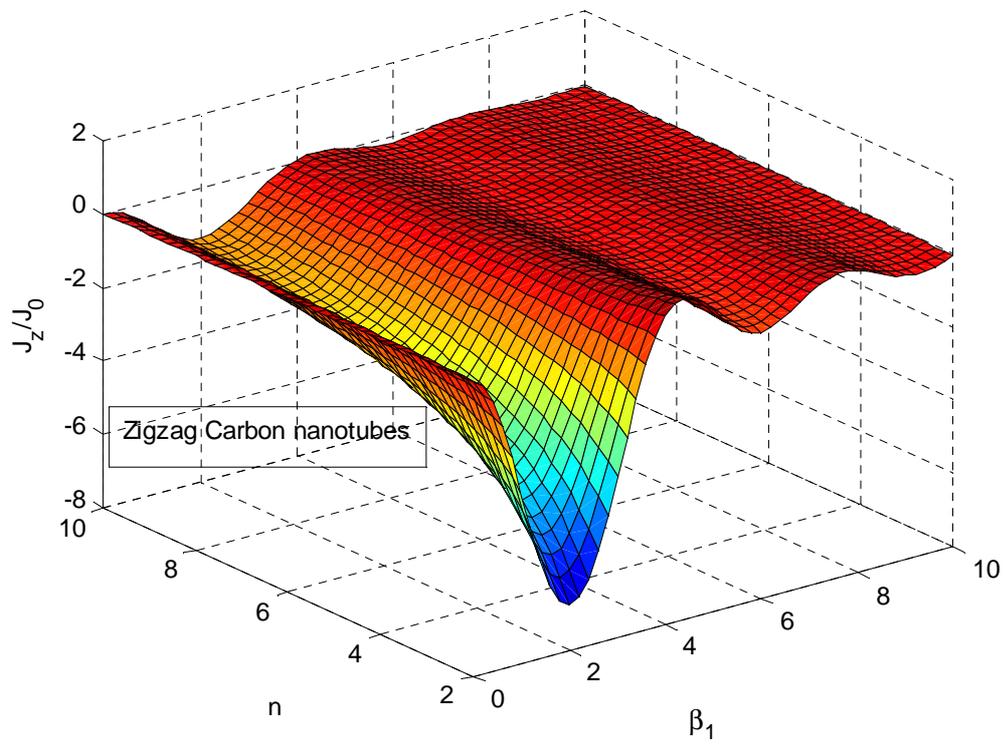

**Fig.4.** $J_z/J_0$ is plotted against $\beta_1$ for Zigzag Carbon nanotubes.

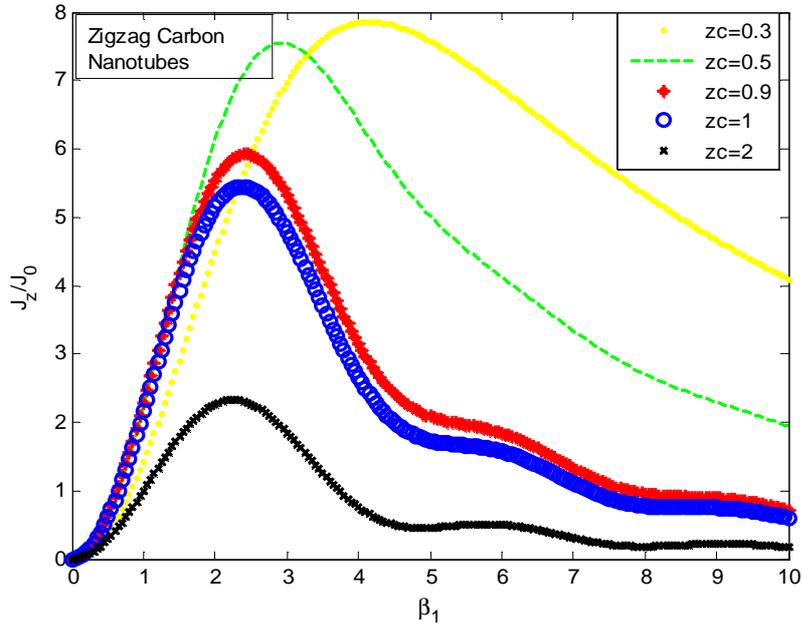

**Fig. 5.** $J_z/J_0$ is plotted against $\beta_1$ for (· — ·) $zc = \Omega\tau = 0.3$; (···) $zc = \Omega\tau = 0.5$; (∗∗∗) $zc = \Omega\tau = 0.9$ (ᵒᵒᵒ) $zc = \Omega\tau = 1$; (···) $zc = \Omega\tau = 2$. When the phase shift $\varphi$ lies between $\frac{\pi}{2}$ and $\frac{3\pi}{2}$